\documentclass[useAMS,usenatbib]{mn2e}

\usepackage{hyperref} 
\usepackage{amsmath} 
\usepackage{amssymb} 
\usepackage{graphicx} 
\usepackage{indentfirst} 
\usepackage{pdflscape} 
\usepackage{placeins} 
\usepackage{nicefrac} 
\usepackage{afterpage} 
\usepackage{rotating} 
\usepackage{booktabs} 
\usepackage{rotating} 
\usepackage{color} 
\newcommand{\kms}{${\rm km\,s}^{-1}$} 
 
\newcommand{\lya}{\mbox{${\rm Ly}\alpha$}} 
\newcommand{\lyb}{\mbox{${\rm Ly}\beta$}} 
 
\newcommand{\PG}{PG\,$1522$+$101$}

\title[]{Discovery of a transparent sightline at $\rho \lesssim 20$\,kpc from an interacting pair of galaxies} 
\author[]{Sean D. Johnson$^{1}$\thanks{E-mail: seanjohnson@uchicago.edu}, Hsiao-Wen Chen$^{1}$\thanks{E-mail: hchen@oddjob.uchicago.edu}, John S. Mulchaey$^{2}$, Todd M. Tripp$^{3}$, 
\newauthor J. Xavier Prochaska$^{4}$, \& Jessica K. Werk$^{4}$\\
$^{1}$Department of Astronomy \& Astrophysics and Kavli Institute for Cosmological Physics, The University of Chicago, Chicago, IL 60637, USA\\
$^{2}$The Observatories of the Carnegie Institute of Washington, 813 Santa Barbara Street, Pasadena, CA 91101, USA\\
$^{3}$Department of Astronomy, University of Massachusetts, Amherst, MA \\
$^{4}$UCO/Lick Observatory, University of California, Santa Cruz, CA} 
\begin{document}

\date{\today}

\pagerange{
\pageref{firstpage}--
\pageref{lastpage}} \pubyear{2013}

\maketitle

\label{firstpage}
\begin{abstract}
	We report the discovery of a transparent sightline at projected distances of $\rho \lesssim 20$ kpc to an interacting pair of mature galaxies at $z=0.12$. The sightline of the UV-bright quasar \PG\, at $z_{\rm em}=1.328$ passes at $\rho = 11.5$ kpc from the higher-mass galaxy ($M_* = 10^{10.6}\,M_\odot$) and $\rho=20.4$ kpc from the lower-mass one ($M_* = 10^{10.0}\,M_\odot$). The two galaxies are separated by $9$ kpc in projected distance and $30$ \kms\, in line-of-sight velocity. Deep optical images reveal tidal features indicative of close interactions. Despite the small projected distances, the quasar sightline shows little absorption associated with the galaxy pair with a total H\,I column density {\it no greater than} $\log \, N({\rm H\,I})/{\rm cm}^{-2}=13.65$. This limiting H\,I column density is already two orders-of-magnitude less than what is expected from previous halo gas studies. In addition, we detect no heavy-element absorption features associated with the galaxy pair with $3$-$\sigma$ limits of $\log\,N({\rm Mg\,II})/{\rm cm}^{-2} < 12.2$ and $\log\,N({\rm O\,VI})/{\rm cm}^{-2} < 13.7$. The probability of seeing such little absorption in a sightline passing at a small projected distance from two non-interacting galaxies is $0.2$\%. The absence of strong absorbers near the close galaxy pair suggests that the cool gas reservoirs of the galaxies have been significantly depleted by the galaxy interaction. These observations therefore underscore the potential impact of galaxy interactions on the gaseous halos around galaxies.
\end{abstract}
\begin{keywords}
	galaxies: haloes -- galaxies: interactions -- quasars: absorption lines 
\end{keywords}

\section{Introduction} \label{section:introduction}

The origin of the division of galaxies into the star-forming galaxies in the blue cloud and the more passive ones of the red sequence in color-magnitude diagrams, along with the corresponding division in morphologies, remains one of the most fundamental questions in galaxy evolution \citep[e.g.][]{Hubble:1936, DeVaucouleurs:1961, Strateva:2001, Bell:2004, Faber:2007}. Galaxies grow by a combination of accreting material from the intergalactic medium (IGM) and mergers \citep[e.g.][]{Genel:2008}, but the build-up of the red sequence requires mechanisms that rapidly halt or ``quench'' star-formation for long time-scales. Possible quenching mechanisms include: shock-heating of accreted gas \citep[e.g.][]{Dekel:2006}, gas stripping due to interactions with a cluster environment \citep[e.g.][]{Moore:1996, Moore:1998, Tonnesen:2009}, and winds driven by starbursts or active galactic nuclei \citep[e.g.][]{Springel:2005}. One particularly intriguing picture is one in which galaxy mergers transform spirals into ellipticals \citep[e.g.][]{Larson:1980, Barnes:1989, Schweizer:1992, Cox:2006, Hopkins:2009}, drive the observed correlations between the passive fraction and galaxy density \citep[e.g.][]{Dressler:1980, Poggianti:2006}, and quench star-formation by stripping gas through galaxy collisions \citep[e.g.][]{Spitzer:1951, Toomre:1972, Balogh:2000}, or ram-pressure stripping \citep[e.g.][]{Gunn:1972, Kenney:2004}. Yet despite decades of study, it is not clear whether galaxies are quenched by interaction related processes or intrinsic mechanisms such as a transition from cold-mode to hot-mode accretion.

Proposed quenching mechanisms operate by heating \citep[e.g.][]{Dekel:2009}, ejecting \citep[e.g.][]{Keres:2009}, or stripping \citep[in cluster environments; e.g.][]{Balogh:2000} the gaseous halos of galaxies, and consequently, the circumgalactic medium (CGM) provides a sensitive laboratory for studying the physical mechanisms that govern galaxy evolution. The diffuse gas of the CGM is generally too low in density to be directly observed in emission with existing facilities, but it can be studied through the use of rest-frame ultraviolet (UV) absorption features in the spectra of background sources. Over the last two decades, a great deal of effort has gone into constraining the properties of the CGM as a function of galaxy mass, luminosity, morphology, and color through observations of absorption features in the spectra of quasars at small projected distances from galaxies (e.g. Bergeron \& Boiss\'{e} 1991; Morris et al. 1993; Lanzetta et al. 1995; Churchill et al. 1996; Chen et al. 1998, 2001a; Tripp et al. 1998; Chen et al. 2001b; Bowen et al. 2002; Simcoe et al. 2006; Stocke et al. 2006; Wakker \& Savage 2009; Chen \& Mulchaey 2009; Wakker \& Savage 2009; Chen et al. 2010; Gauthier et al. 2010, Steidel et al. 2010, Lovegrove \& Simcoe 2011; Prochaska et al. 2011; Tumlinson et al. 2011; Thom et al. 2012; Werk et al. 2013). \nocite{Bergeron:1991, Morris:1993, Lanzetta:1995, Churchill:1996, Chen:1998, Tripp:1998, Chen:2001a, Chen:2001b, Bowen:2002, Simcoe:2006, Stocke:2006, Chen:2009, Wakker:2009, Chen:2010, Gauthier:2010, Steidel:2010, Lovegrove:2011, Prochaska:2011, Tumlinson:2011, Thom:2012, Werk:2013} These studies show that the bimodal distribution of galaxies in color-magnitude diagrams is reflected in the gas content of the CGM. Observations of both H\,I components and heavy-element species such as O\,VI and Mg\,II demonstrate that cool ($T < 10^{6}$ K) gas is common around passive galaxies but with reduced covering fraction relative to star-forming galaxies \citep[][]{Chen:2009, Chen:2010, Gauthier:2010, Tumlinson:2011, Thom:2012}. The mechanisms responsible for quenching star-formation in the red sequence must remove or transform the highly ionized gaseous halos of quenched galaxies while maintaining the lower ionization-state gas but at reduced levels.

The CGM is likely to be affected by galaxy interactions such as ram-pressure stripping and tidal forces, and studies of the impact of galaxy environment on the CGM may help constrain the relationship between quenching and environmental factors. Nevertheless, while existing studies of the CGM explore a range of galaxy properties, the influence of group and cluster environments on the CGM remains largely unexplored. Group galaxies exhibit more extended Mg\,II halos than isolated galaxies \citep[][]{Chen:2010, Bordoloi:2011}, and galaxy clusters exhibit an increased incidence of strong Mg\,II absorbers at $\rho < 1$ Mpc relative to the field \citep[][]{Lopez:2008}. At the same time, members of the Virgo cluster show a reduced covering fraction in \lya\, absorption \citep[]{Yoon:2012}. Additional studies that compare the CGM of cluster, group, and field galaxies are warranted.

In this paper, we report the discovery of a transparent sightline harboring little $T < 10^6$ K gas at a small projected distance to an interacting pair of mature galaxies at $z=0.12$. The sightline of the UV-bright quasar \PG\, at $z_{\rm em}=1.328$ passes at a projected distance of $\rho \lesssim 20$ kpc from the galaxy pair (see the right panel of Figure \ref{figure:galaxies}), enabling a study of the gas at small projected distances from the interacting pair. The galaxies are separated by $9$ kpc in projected distance. We refer to the more luminous pair member as G1 and to the less luminous member as G2 throughout the paper. This discovery represents an extreme example demonstrating the potential impact of galaxy environment on the CGM.

This paper proceeds as follows. In Section \ref{section:data} we present the imaging and spectroscopy of the galaxy pair as well as the UV and optical spectroscopy of \PG. In Section \ref{section:galaxy}, we discuss the properties of the galaxy pair including: absolute magnitudes, colors, star-formation rates, gas-phase metallicities, and ages of the dominant stellar populations. In Section \ref{section:absorption} we review the absorption features revealed in the quasar sightline and place limits on the possible presence of key ions. Finally, in Section \ref{section:discussion}, we briefly discuss the implications of our study. Throughout the paper, we adopt a $\Lambda$ cosmology with $\Omega_{\rm m}=0.3$, $\Omega_{\Lambda} = 0.7$, and $H_0=70$ \kms\,${\rm Mpc}^{-1}$. We perform K corrections with the IDL kcorrect library described in \cite{Blanton:2007}. All magnitudes reported here are in the AB system.

\section{Data} \label{section:data} \PG\, is a UV bright quasar at high redshift ($z_{\rm em} =1.328$), and consequently it represents a prime target for absorption studies of the IGM. A wealth of imaging and spectroscopic data in the \PG\, field are available including: (1) optical and near-infrared imaging of the field, (2) optical spectroscopy of the galaxy pair, and (3) optical and UV spectroscopy of the quasar. In this section, we describe each of these three datasets and the related data-reduction and processing steps. The imaging data are summarized in Table \ref{table:imaging_observations}, and the spectroscopic data are summarized in Table \ref{table:spectroscopic_observations}.

\begin{table}
	\caption{Summary of Imaging Observations} \label{table:imaging_observations} \centering 
	\begin{tabular}
		{ccccccccc} \hline \hline & & Exposure & FWHM & \\
		\multicolumn{1}{c}{Instrument} & Filter & Time (s) & ($''$) &Date \\
		\hline VIMOS & $B$ & $470$ & $0.8$ &2004 May \\
		VIMOS & $V$ & $350$ & $0.7$ &2004 May\\
		VIMOS & $R$ & $90$ & $0.7$ &2004 May\\
		VIMOS & $I$ & $180$ & $0.6$ &2004 May\\
		UKIDSS & $J$ & $40$ & $1.1$ &2007 Mar \\
		UKIDSS & $H$ & $40$ & $0.9$ &2007 Mar\\
		UKIDSS & $K$ & $40$ & $0.8$ &2007 Mar \\
		LDSS3 & $r$ & $1500$ & $0.7$ &2009 Apr \\
		\hline 
	\end{tabular}
\end{table}
\begin{table}
	\caption{Summary of Spectroscopic Observations} \label{table:spectroscopic_observations} \centering \resizebox{\columnwidth}{!}{ 
	\begin{tabular}
		{ccccccccc} \hline \hline &Instrument\,/ & Exposure & FWHM & \\
		\multicolumn{1}{c}{Target}& Disperser & Time (ks) & (\AA) & Date \\
		\hline \PG & HIRESb & $1.6$ & $0.06$ & 2010 Mar \\
		\PG & COS\,/\,G130M &$16$ & $0.07$ & 2010 Sep \\
		\PG & COS\,/\,G160M & $23$ & $0.07$ & 2010 Sep\\
		G$1$ & IMACS\,/\,$200$l &$14$ & $9$ & 2009 May\\
		G$1$ & IMACS\,/\,$200$l & $11$ & $9$ & 2010 May\\
		G$2$ & IMACS\,/\,$200$l & $11$ & $9$ & 2010 May\\
		\hline 
	\end{tabular}
	} 
\end{table}

\subsection{Galaxy imaging} \label{section:data_gal} 
\begin{figure*}
	\centering 
	\includegraphics[scale=0.495]{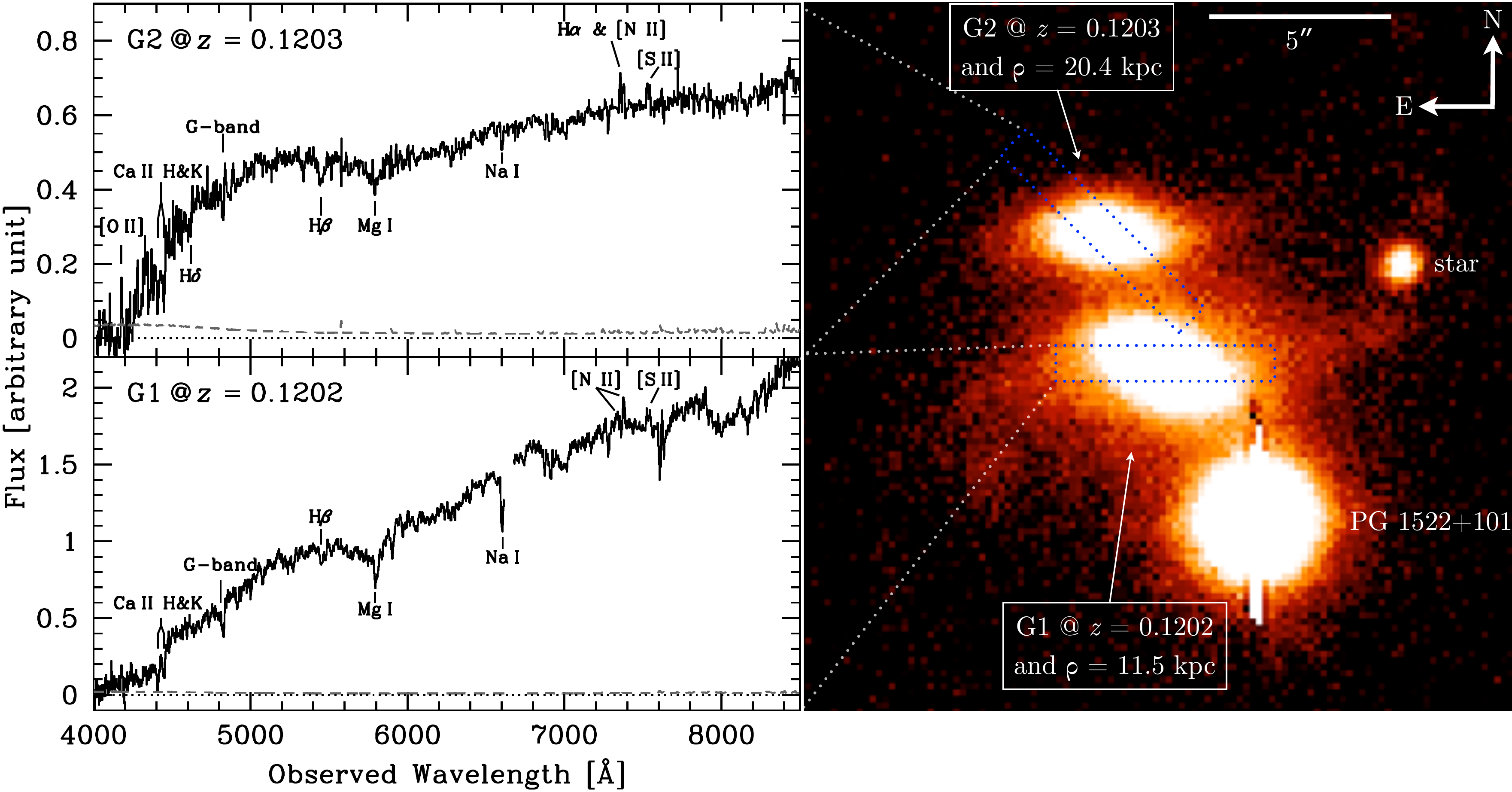} \caption{{\it Left}: IMACS spectra of the galaxy pair with the brighter member, G1, in the {\it bottom} panel and the fainter member, G2, in the {\it top} panel. The spectra are shown in black histogram, and the $1$-$\sigma$ error arrays are shown in gray dashed line. The zero-flux level is shown in black dotted line. Key spectral features in the galaxy spectra are labelled. Note that the flux calibrations of G1 and G2 are relative, so the continuum slopes of the two galaxies should not be compared. {\it Right}: LDSS3 $r$-band image of the galaxy pair and the quasar PG\,$1522$+$101$, scaled and stretched to reveal the presence faint tidal features. The faint tidal features are seen most prominently in the vicinity of G1. Objects in the field are labelled in white and the spatial scale and orientation of the image are shown in the top right corner. The slit positions and orientations used to observe G1 and G2 spectroscopically are shown in blue dotted line and white dotted lines lead the the eye from the slits to the corresponding spectra shown on the left. The star labelled in the image has been spectroscopically confirmed as part of our redshift survey in the quasar field.} \label{figure:galaxies} 
\end{figure*}

Optical $r$-band images of the field around \PG\, were obtained using the Low Dispersion Spectrograph 3 (LDSS3) on the Magellan Clay Telescope in April of 2009 as pre-imaging for our blind redshift survey in the fields of UV bright quasars \citep[see][]{Chen:2009}. The observations consist of a sequence of five exposures of 300 seconds duration each with a full-width at half maximum (FWHM) seeing of ${\rm FWHM}=0.7''$. The LDSS3 image of the field was reduced using standard IRAF routines and is displayed in the right panel of Figure \ref{figure:galaxies}. The depth of the LDSS3 image reveals the presence of faint tidal features around the galaxy pair.

In addition to our LDSS3 images, multi-band optical and near-infrared images of the field are available in the public European Southern Observatory (ESO) Archive and in the United Kingdom Infrared Deep Sky Survey \citep[UKIDSS;][]{Hewett:2006, Casali:2007, Lawrence:2007, Hambly:2008, Hodgkin:2009}. The ESO imaging of the field (PI: Christopher Mullis, PID: 71.A-3030) was acquired serendipitously with the Visible MultiObject Spectrograph \citep[VIMOS;][]{LeFevre:2003} in May of 2004 as part of an unrelated program. The VIMOS imaging of the pair includes a single exposure in each of the $B$, $V$, $R$, and $I$ filters for $470$, $350$, $90$, and $180$ seconds respectively, under seeing conditions ranging from ${\rm FWHM}=0.8''-0.6''$. We reduced the VIMOS imaging with a custom set of software that bias subtracted and flat-fielded the data and in the case of the $I$-band image, removed the fringe pattern. We determined the zero-points of the VIMOS imaging data using field stars observed by the Sloan Digital Sky Survey \citep[SDSS;][]{York:2000} and the SDSS-to-Johnson filter conversion formulae derived by Robert Lupton in 2005 and given in the SDSS Data Release 9 \citep[][]{Ahn:2012} documentation. The UKIDSS imaging of the field includes $40$ second exposures in the $J$, $H$, and $K$ filters taken under seeing conditions of ${\rm FWHM} = 1.1''$, $0.9''$, and $0.8''$ respectively. We retrieved the UKIDSS images of the field from the public UKIDSS Data Release 8 website and combined them into a single frame covering the full VIMOS field of view in each filter using the SWARP tool \citep{Bertin:2002}. We converted the UKIDSS photometric zeropoints to the AB system using the conversions in \cite{Hodgkin:2009}.

In order to extract galaxy photometry from the VIMOS and UKIDSS data, we registered the images to a common origin and pixel scale using standard IRAF routines. Next, we formed a detection image by summing the background subtracted VIMOS $V$, $R$, and $I$-band images to enable detection of the low surface-brightness extent of the galaxies. We then defined the isophotal apertures ($41$ and $24$ square arc seconds in area for G1 and G2) of the two galaxies in the summed detection image and calculated the total flux in these apertures in each bandpass using SExtractor \citep{Bertin:1996} in double-image mode. Finally, we corrected the galaxy photometry for foreground Milky Way extinction following \cite{Schlegel:1998}. The observed magnitudes of the two galaxies and other photometric properties are shown in Table \ref{table:photometric}.

\begin{table*}
	\caption{Summary of Galaxy Photometric Properties} \label{table:photometric} \resizebox{
	\textwidth}{!}{ 
	\begin{tabular}
		{cccccccccccccccc} \hline & $\theta$ & $\rho$ & ${\alpha}^{\rm a}$ & ${i}^{\rm b}$ & \multicolumn{7}{c}{Apparent, Isophotal Magnitudes} & \multicolumn{3}{c}{Absolute Magnitudes$^{\rm c}$} \\
		\cmidrule(lr){6-12} \cmidrule(lr){13-15} ID&($''$)&$($kpc$)$&$($deg$)$&$($deg$)$& $B_{AB}$ & $V_{AB}$ & $R_{AB}$ & $I_{AB}$ & $J_{AB}$ & $H_{AB}$ & $K_{AB}$ & $M_B$ & $M_R$ & $M_K$ & ${\log\,M_*/M_\odot}^{\rm c}$ \\
		\hline G1 & $5.1$ & $11.5$ & $32$ & $56$ & $19.58\pm0.03$ & $18.32\pm0.04$ & $17.75\pm0.07$ & $17.18\pm0.06$ & $16.74\pm0.06$ & $16.38\pm0.06$ & $16.56\pm0.06$ & $-19.81$ & $-21.12$ & $-21.93$ & $10.6$ \\
		G2 & $9.4$ & $20.4$ & $56$ & $58$ & $20.40\pm0.03$ & $19.39\pm0.04$ & $18.91\pm0.07$ & $18.49\pm0.06$ & $18.38\pm0.12$ & $17.92\pm0.12$ & $17.98\pm0.11$ & $-18.82$ & $-19.93$ & $-20.50$ & $10.0$ \\
		\hline \multicolumn{16}{l}{\bf Notes} \\
		\multicolumn{16}{l}{$^\mathrm{a}$ The azimuthal angle of the major axis of the galaxy from the line connecting the galaxy to the quasar sightline, measured North through East. } \\
		\multicolumn{16}{l}{$^\mathrm{b}$ Inclination angle of the galaxy disk measured from the major-to-minor axis ratio assuming an intrinsic, edge-on major-to-minor axis ratio of $0.2$.} \\
		\multicolumn{16}{l}{$^\mathrm{c}$ Absolute, rest-frame magnitudes and stellar mass determined using the IDL kcorrect library v4\_2 \citep{Blanton:2007} which uses templates created with the} \\
		\multicolumn{16}{l}{\,\,\,\, \cite{Bruzual:2003} stellar population synthesis code and emission-line models from \cite{Kewley:2001}.} \\
	\end{tabular}
	} 
\end{table*}
\begin{table*}
	\caption{Summary of Galaxy Spectroscopic Properties} \centering \label{table:spectroscopic} \resizebox{
	\textwidth}{!}{ 
	\begin{tabular}
		{cccccccccccc} \hline \hline & & \multicolumn{5}{c}{absorption lines, $W_r\,\,$(\AA)} & \multicolumn{4}{c}{emission lines, $W_r\,\,$(\AA)} & ${{{\rm SFR}_{{\rm H}\alpha}}}^{\rm c}$ \\
		\cmidrule(lr){3-7} \cmidrule(lr){8-11} ID & $z$ & Ca\,II\,\,$\lambda3934$ & G-band & Fe\,I\,\,$\lambda4384$ & Mg\,I\,\,$\lambda5185$ & Na\,I\,\,$\lambda5897$ & [O\,I]$\lambda 6302$ & ${{\rm H}\alpha}^{\rm a}$ & ${\rm [N\,II]}\,\,\lambda6585$ & ${12 + \log\,{\rm O/H}}^{\rm b}$ & ${(M_\odot\,{\rm yr}^{-1})}$ \\
		\hline G1 & $0.1202 $ & $10.0\pm0.7$ & $5.3\pm0.2$ & $1.7\pm0.3$ & $5.1\pm0.2$ & $3\pm1$ & $<0.1$ & $-1.9\pm0.3$ & $-1.1\pm0.2$ & $\approx 9.0$ & $\approx 0.08$\\
		G2 &$0.1203$ & $6.9\pm1.3$ & $4.3\pm0.4$ & $1.3\pm0.3$ & $2.4\pm0.3$ & $1.7\pm0.2$ & $<0.2$& $-4.0\pm0.3$ & $-1.4\pm0.2$ & $\approx8.7$ & $\approx0.06$\\
		\hline \multicolumn{12}{l}{\bf Notes} \\
		\multicolumn{12}{l}{$^\mathrm{a}$ H$\alpha$ emission-line equivalent width corrected for underlying stellar absorption based on stellar population synthesis models.} \\
		\multicolumn{12}{l}{$^\mathrm{b}$ Gas-phase Oxygen abundance estimate from the ``N2'' index, ${\rm N2} = \log\,([{\rm N\,II}]\,\lambda 6585/{\rm H}\alpha)$, using the relationship from \cite{Pettini:2004}.} \\
		\multicolumn{12}{l}{\,\,\,\, For reference, the estimated Oxygen abundance of the Sun is $12 + \log\,{\rm O/H}=8.7$ \citep[][]{Prieto:2001}.} \\
		\multicolumn{12}{l}{$^\mathrm{c}$ The star-formation rate measurements for G1 and G2 is derived from H$\alpha$. The H$\alpha$ line luminosity is estimated from the absolute, rest-frame $R$-band} \\
		\multicolumn{12}{l}{\,\,\,\, magnitude and the H$\alpha$ equivalent width observed in our IMACS spectra. The H$\alpha$ line luminosity is then converted to a star-formation rate using the} \\
		\multicolumn{11}{l}{\,\,\,\, relation from \cite{Kennicutt:2012}.} \\
	\end{tabular}
	} 
\end{table*}

\subsection{Galaxy spectroscopy} Galaxies G1 and G2 were observed spectroscopically as part of our blind redshift survey in the fields of UV bright quasars with the short camera of the Inamori-Magellan Areal Camera and Spectrograph \citep[IMACS;][]{Dressler:2011} in multi-object mode with the $200$l grating and $1''$ slitlets. In this configuration, IMACS delivers spectral coverage from $\approx 4000$ to $\approx9000$\,\AA\, with spectral resolution of ${\rm FWHM}\approx 9\,$\AA. G1 was observed in May of 2009 for a total of $14\,{\rm ks}$ and again in May of 2010 for a total of $11\,{\rm ks}$. G2 was observed for a total of $11\,{\rm ks}$ in May of 2010. The IMACS data were reduced using the Carnegie Observatories System for MultiObject Spectroscopy\footnote{ \href{http://code.obs.carnegiescience.edu/cosmos}{http://code.obs.carnegiescience.edu/cosmos}} (COSMOS) version 2.16 as described in \cite{Chen:2009}. The spectra were flux calibrated in a relative sense using a bright star on the multi-object mask, but no absolute flux standard is available. Since G1 and G2 were observed on different masks and were flux calibrated using different stars, the measured spectral slope of the galaxies cannot be compared. The galaxy redshifts are determined by fitting SDSS galaxy eigenspectra \citep[e.g.][]{Yip:2004} and are corrected for the heliocentric motion of the Earth using the IRAF rvcorrect routine. The final redshifts for G1 and G2 are $z=0.1202$ and $z=0.1203$ respectively. Based on a comparison of galaxy redshifts in the \PG\, field measured by both our survey and the SDSS\footnote{ We note that G1 and G2 were not observed spectroscopically by the SDSS.}, we estimate that the $1$-$\sigma$ uncertainty in the redshifts corresponds to $\approx 60$ \kms. The reduced spectra of galaxies G1 and G2 are shown in the bottom left and top left panels of Figure \ref{figure:galaxies} respectively, and the spectroscopic properties of the two galaxies are shown in Table \ref{table:spectroscopic}.

\subsection{COS and HIRES quasar spectroscopy} \label{section:data_quasar}

High quality UV spectra of 
\PG\, were acquired with the Cosmic Origins Spectrograph \citep[COS;][]{Green:2012} aboard the {\it Hubble Space Telescope} ({\it HST}) as part of a blind survey for IGM absorption (PI: Todd Tripp, PID: $11741$). The {\it HST} observations include six exposures totaling $16\,{\rm ks}$ with the G130M grating (covering the spectral range $1150-1450$\,\AA\,with a spectral resolution of $\rm{FWHM} \approx 16$ \kms) and eight exposures totaling $23\,{\rm ks}$ with the G160M grating ($1400-1800$\,\AA\,, ${\rm FWHM} \approx 16$ \kms). We combined the 1D, calibrated COS spectra using a custom software suite first described in \cite{Meiring:2011}. COS is a photon-counting instrument with very low dark backgrounds and minimal scattered light \citep[see ][]{Green:2012}. This has two implications: (1) it is straightforward to estimate flux uncertainties based on photon counting statistics, and (2) in pixels with low flux levels (e.g., pixels in the cores of deep absorption lines), total counts in an exposure can be very low resulting in asymmetric (Poissonian) flux uncertainties. We evaluated these uncertainties by coadding exposures and accumulating gross and background counts in each pixel, which were used to determine uncertainties based on Poisson statistics \citep[e.g.][]{Gehrels:1986}. To correct for inaccuracies in the COS wavelength calibration, we aligned the exposures by cross correlating well-detected interstellar or extragalactic absorption lines, and likewise the G130M and G160M segments were aligned by cross correlating lines of comparable strength. COS spectra also exhibit fixed-pattern noise, due to shadows from the photocathode grid wires. We corrected the data for these fixed-pattern features using flatfields derived from high-S/N white dwarf observations (D. Massa, private communication) filtered to remove spurious high-frequency noise. Fixed pattern noise due to the hexagonal pattern in the micro-channel
plate cannot be fully removed, but the significance of this fixed pattern noise is reduced by the
combination of multiple exposures acquired with different central-wavelength settings.

In addition to the UV COS spectrum, high-resolution optical spectra of the quasar were acquired with the High Resolution Echelle Spectrometer with settings optimized for short-wavelength observations \citep[HIRESb;][]{Vogt:1994} on the Keck\,I telescope (PI: Jason Prochaska, PID: U066Hb) in March of 2010. The HIRESb observations include two exposures totaling $1.6\,{\rm ks}$ with the $0.86''$ slit (covering the spectral range $3000-6000$\,\AA\, with spectral resolution of ${\rm FWHM}\approx6$\,\kms). We reduced the HIRES spectrum with standard techniques using the HIRedux package\footnote{ \href{http://www.ucolick.org/~xavier/HIRedux/index.html}{http://www.ucolick.org/$\sim$xavier/HIRedux/index.html}} bundled in the XIDL library. 

\section{Galaxy pair properties} \label{section:galaxy}

Galaxies G1 and G2 are separated from each other by $8.95\,$ kpc in projected distance and $30 \pm 60$ \kms\, in line-of-sight velocity. The galaxies exhibit low-surface brightness features (see the right panel of Figure \ref{figure:galaxies}) characteristic of tidal disruption (though we cannot rule out the possibility of faint spiral arms). These faint features combined with the small projected distance and radial velocity difference indicate that G1 and G2 are strongly interacting.

G1 has a rest-frame $R$-band absolute magnitude $M_R = -21.12$, rest-frame optical color $M_B - M_R = 1.3$, and a best-fit stellar mass\footnote{ Best-fit stellar masses are an ancillary product of the kcorrect code.} of $\log\,M_*/M_\odot \approx 10.6$, while G2 has $M_R = -19.93$, $M_B - M_R = 1.1$, and $\log\,M_*/M_\odot \approx10.0$. Adopting the stellar-to-halo mass relation of \cite{Moster:2010} from an abundance matching approach, we estimate that the halo mass of the interacting galaxy pair is $\log\,M_{\rm halo}/M_\odot \approx 12.1$ with a corresponding virial velocity $v_{\rm vir} \approx 200$ \kms.

The colors of both galaxies are consistent with those of an evolved stellar population with some trace of ongoing star-formation. We estimate the star-formation rates of G1 and G2 using a combination of our imaging and spectroscopic data. First, we measure the rest-frame, H$\alpha$ emission equivalent widths observed in the galaxy spectra. Our IMACS spectrum of G1 shows only weak H$\alpha$ emission despite the presence of both [N\,II] and [S\,II] emission due to underlying stellar absorption. We correct for the stellar absorption based on the best-fit stellar-population synthesis (SPS) model described below after convolving and rebinning the SPS model spectrum to match the IMACS resolution and wavelength grid. The stellar absorption corrected H$\alpha$ emission equivalent width of G1 is $W_r({\rm H}\alpha) = -1.9\pm0.3$\,\AA. Similarly, we measure a stellar absorption corrected equivalent width of $W_r({\rm H}\alpha)=-4.0\pm0.3$\,\AA\, for G2. Next, we estimate the H$\alpha$ line luminosities of G1 and G2 using these equivalent widths and the $R$-band absolute magnitudes of the galaxies. Finally, we convert the H$\alpha$ luminosities to star-formation rates using the formula from \cite{Kennicutt:2012}. The estimated star-formation rates of G1 and G2 are ${\rm SFR}_{{\rm H}\alpha}\approx0.08\,M_\odot\,{\rm yr}^{-1}$ and ${\rm SFR}_{{\rm H}\alpha}\approx 0.06\,M_\odot\,{\rm yr}^{-1}$ respectively.

In addition, we measure the [N\,II] $\lambda 6585$ emission equivalent widths of G1 and G2 in order to estimate the gas-phase Oxygen abundance of the two galaxies using the ``N2'' index from \cite{Pettini:2004}. Galaxies G1 and G2 exhibit [N\,II] equivalent widths of $W_r({\rm N\,II\,\lambda6585})=-1.1\pm0.2$ and $-1.4\pm0.2$ respectively. The corresponding Oxygen abundance estimates for G1 and G2 are $12 + \log\,({\rm O/H}) \approx 9.0$ and $12 + \log\,({\rm O/H}) \approx 8.7$. We note that our IMACS spectra are dominated by the central regions of these two galaxies, and in the presence of metallicity gradients, our gas-phase Oxygen abundances are upper limits. For reference, the estimated Oxygen abundance of the Sun is $12 + \log\,{\rm O/H}=8.7$ \citep[][]{Prieto:2001}.
\begin{figure*}
	\centering 
	\includegraphics[scale=0.89]{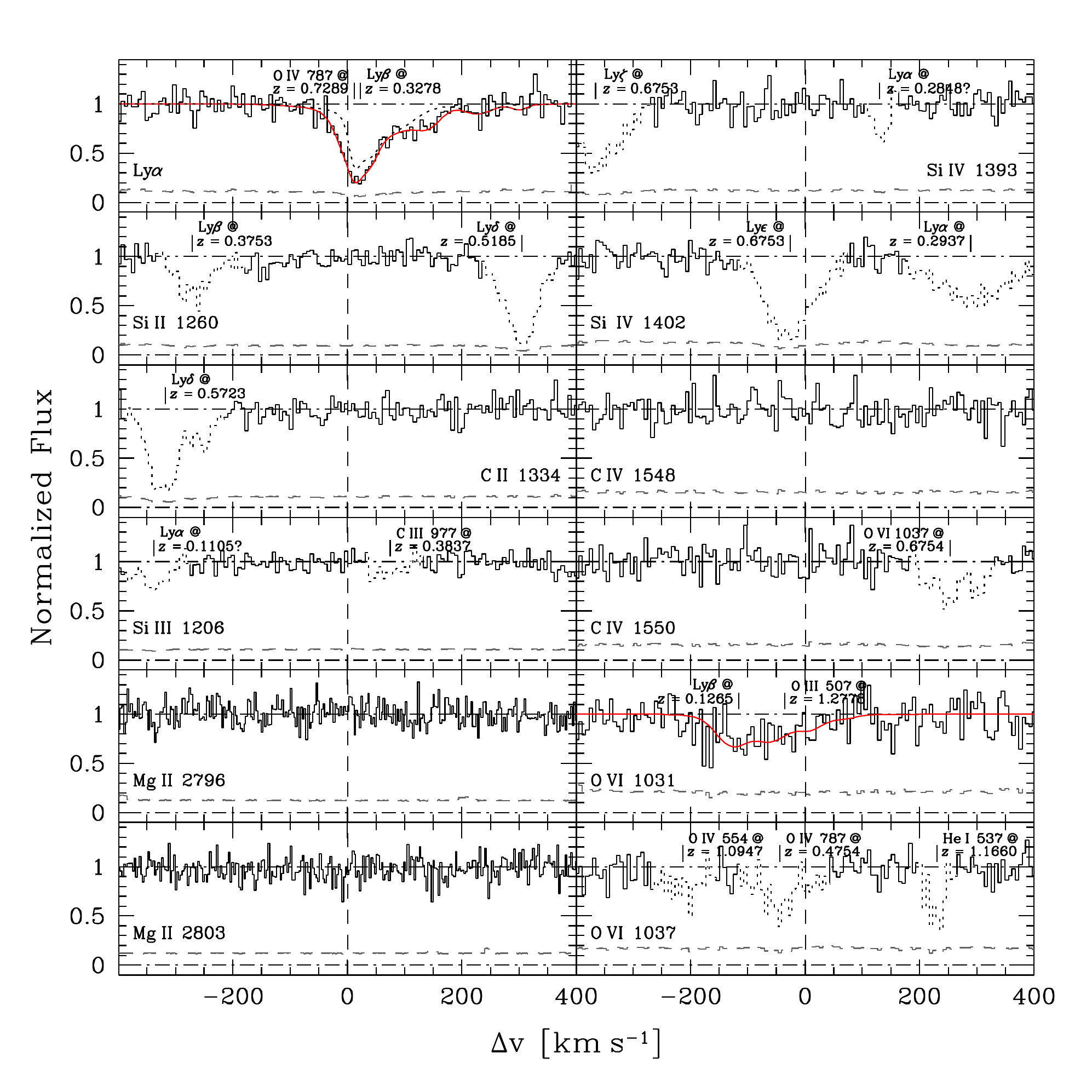}
	\caption{Absorption profiles of halo gas around the interacting galaxy pair. Low-ionization species are shown on the {\it left} and high ionization species on the {\it right}. Zero velocity corresponds to the redshift of G1 at $z=0.1202$. In each panel, the continuum normalized spectrum is shown in solid black histogram, and the $1$-$\sigma$ error array is shown in gray dashed line. Contaminating features are dotted out and labelled except in the case of \lya\, and O\,VI $\lambda 1031$ which are discussed in detail both here and in the main text.
The \lya\, absorption associated with the galaxy pair is contaminated by both \lyb\,  at $z = 0.3278$ and O\,IV
at $z=0.7289$.
We were able to accurately determine the strength of the contaminating \lyb\, by
simultaneously fitting a Voigt profile to the associated \lya\, and ${\rm Ly}\gamma$ features at $z=0.3278$.  The strength of O\,IV at $z=0.7289$ is, however, more difficult to estimate given the multiphase nature of the gas \citep[see][]{Lehner:2013}.  We present in the top-left panel the observed absorption in black histogram
together with the model for the contaminating \lyb\, absorption in black dotted line.
The data exhibit excess absorption which could be due to
\lya\, associated with the galaxy pair or O\,IV at $z=0.7289$.
Given the uncertain strength of O\,IV at $z=0.7289$,
we place a limit on the possible presence of H\,I
associated with the galaxy pair by attributing all of the observed
absorption to \lya\, at $z=0.12$.
After removal of the \lyb\,  at $z = 0.3278$ (see the top panel of Figure \ref{figure:absorption1}),
the data are best fit by two Voigt components at 
$\Delta v = +7$ and $+140$ \kms\,
with column densities of $\log N({\rm H\,I})/{\rm cm}^{-2}=13.5\pm0.06,13.0\pm0.07$
and line widths of $b=28\pm4,28\pm7$ \kms\, respectively.
The best-fit model including these two components and the contaminating \lyb\, is shown in solid red line.
The O\,VI absorption associated with G1 and G2 is contaminated
by \lyb\, at $z=0.1265$ and O\,III $\lambda 507$ at $z=1.2776$ as shown in the second from the bottom panel on the right.
In order to model the contaminating absorption,
we perform a Voigt profile analysis of the \lya\, absorption at $z=0.1265$
and O\,III $\lambda 702$ absorption at $z=1.2776$.
The best-fit model is shown in red line.
After removal of the contaminating absorption
(see the bottom panel of Figure \ref{figure:absorption1}), no excess absorption is observed, and we place a $3$-$\sigma$
limit of $\log\,N\rm{(O\,VI)/{\rm cm}^{-2}} < 13.7$
on the O\,VI associated with the galaxy pair.
We do not display the N\,V doublet because it is heavily blended with a Lyman-limit system at $z=0.5185$. } \label{figure:absorption} 
\end{figure*}

The relatively high [N\,II] $\lambda6585$ to H$\alpha$ line ratios observed in G1 and G2 raise the possibility that the galaxies host low-ionization nuclear emission-line regions (LINER) powered by active galactic nuclei (AGN). To address this possibility, we measure $3$-$\sigma$ upper limits on the equivalent width of [O\,I] $\lambda6302$ of $W_{\rm r}({\rm [O\,I]}\,\lambda6302) < 0.1$\,\AA\, for G1 and $W_{\rm r}({\rm [O\,I]}\,\lambda6302) < 0.2$\,\AA\, for G2. These non-detections of [O\,I] emission place limits on the [O\,I] to H$\alpha$ emission-line ratios of $\log\,{\rm [O\,I]}/{\rm H}\alpha < -1.3$. This limit indicates that neither galaxy hosts a LINER AGN based on the models of \cite{Kewley:2006}. In addition, neither galaxy falls in the AGN region of the $W_r({\rm H}\alpha)$ versus $\log\,{\rm N\,II}/{\rm H}\alpha$ classification system developed by \cite{Fernandes:2011} for weak emission-line galaxies. Finally, the images of G1 and G2 show no evidence for nuclear point-sources. We therefore conclude that neither G1 nor G2 hosts an AGN.

We also estimate the ages of the dominant stellar populations of G1 and G2 using a maximum-likelihood analysis that compares Stellar Population Synthesis (SPS) models with the observed properties of the two galaxies. We compute the SPS model spectra using the \cite{Bruzual:2003} code with updated stellar libraries and evolutionary tracks \citep{Bruzual:2010} and the dust model from \cite{Charlot:2000} which is parameterized by the optical depth $\tau_V$ applied to stars younger than $10$ Myr and a fraction $\mu$ of $\tau_V$ in the diffuse interstellar medium (and hence applied to both young and old stars). The model spectra span a range in age from $\log\,t/{\rm Gyr}=-0.2$ to $1.1$ in steps of $\Delta \log\,t/{\rm Gyr} = 0.03$; a range in the $e$-folding time from $\tau=0.1$ to $0.5$ Gyr in steps of $\Delta \tau = 0.1$ Gyr; a range in metallicity covering $Z=0.005, 0.02, 0.2, 0.4, 1.0, $ and $2.5\,Z_\odot$; and a range in dust properties with $\tau_V$ ranging from $\tau_V=0$ to $2$ in steps of $\Delta \tau_V=0.2$ and $\mu=0$ to $1$ in steps of $\Delta \mu = 0.2$.

We compute the likelihood of each SPS model realization by comparing the apparent magnitudes of G1 and G2 and the rest-frame absorption-line equivalent widths of ${\rm Ca\,II}\,\lambda 3934$, the G-band, ${\rm Fe\,I}\,\lambda 4384$, ${\rm Mg\,I}\,\lambda 5897$, and ${\rm Na\,I}\,\lambda 5897$ of the two galaxies measured in our IMACS spectra (see Table \ref{table:spectroscopic}). We include the absorption-line equivalent width measurements in our likelihood analysis in order to help break the age-metallicity degeneracy.
\begin{table*}
	\centering \caption{Summary of absorption limits associated with the interacting galaxy pair at $z=0.1202$.} \label{table:absorption} 
	\begin{tabular}
		{ccccccccccc} \hline \hline & & \multicolumn{2}{c}{H\,I} & Si\,II & C\,II & Si\,III & Mg\,II & Si\,IV & C\,IV & O\,VI \\
		\cmidrule(lr){3-4} \cmidrule(lr){5-11} & $\Delta v$ & & $b$ \\
		$z$ & $($\kms$)$ & $\log\,N/{\rm cm^{-2}}$ & $($\kms$)$ & \multicolumn{7}{c}{$\log\,N/{\rm cm^{-2}}$} \\
		\hline
		
		$0.12022$ & $+7$ & $<13.5 \pm 0.06^{\rm a}$ & $28\pm4$ & $<12.4^{\rm b}$ & $<13.3^{\rm b}$ & $<12.2^{\rm b}$ & $<12.2^{\rm b}$ & $<12.7^{\rm b}$ & $<13.2^{\rm b}$ & $<13.7^{\rm a}$ \\
		$0.12072$ & $+140$ & $<13.0 \pm 0.07^{\rm a}$ & $28\pm7$ & $$ & $$ & $$ & $$ & $$ & $$ & $$ \\
		\hline \multicolumn{11}{l}{\bf Notes} \\
		\multicolumn{11}{l}{$^\mathrm{a}$ Features are contaminated by intervening systems. Limits and measurements are based on a Voigt profile analysis described} \\
		\multicolumn{11}{l}{$$\,\,\,\,\,in Section \ref{section:absorption}.} \\
		\multicolumn{11}{l}{$^\mathrm{b}$ Limits are $3$-$\sigma$ and calculated over a $300$\,\kms\, window.} \\
	\end{tabular}
\end{table*}

The maximum-likelihood age of the dominant stellar population of G1 is $\log\,t/{\rm Gyr}=0.6 \pm 0.1$ and that of G2 is $\log\,t/{\rm Gyr}=0.4^{+0.2}_{-0.1}$ after marginalizing over all other parameters. The stellar mass estimates of the two galaxies from our SPS modeling are consistent with those from kcorrect within uncertainties. The mean values of the other parameters are $Z=1.1, 0.2\,Z_\odot$; $\tau=0.3, 0.3$ Gyr; $\tau_v = 0.5, 0.8$; and $\mu = 0.3,0.4$ for G1, G2. The mean star-formation rates averaged over the last $10$ Myr are ${\rm SFR} = 0.01 \, M_\odot\,{\rm yr}^{-1}$ for both G1 and G2. The specific star-formation rate of G2 is therefore approximately four times that of G1 since G1 is four times more massive. This is consistent with the bluer colors of G2 relative to G1.

No other spectroscopically observed galaxies are known in the proximity of the interacting pair. We carried out a search for galaxies brighter than $r=20$ with SDSS photometric redshifts \citep[][]{Csabai:2007, Carliles:2010} consistent with $z=0.12$ and within $\theta = 2'$ of the quasar sightline. There are three galaxies meeting these criteria at $\theta = 1.05', 1.88',$ and $1.99'$ ($\rho=141, 254, $ and $269$ kpc at $z=0.12$) with apparent $r$-band magnitudes of $r=19.5, 19.8, $ and $18.9$ and photometric redshifts $z_{\rm phot} = 0.17 \pm 0.03, 0.16 \pm 0.03, $ and $0.11 \pm 0.02$. Therefore, G1 and G2 may be part of a small galaxy group.

\section{The transparent sightline} \label{section:absorption} The sightline of \PG\, probes the environment of the interacting galaxy pair at $\rho=11.5$\,kpc and an angle of $32$ degrees from the major axis of G1, and at $\rho=20.4$\,kpc and an angle of $56$ degrees from the major axis of G2. Despite the small impact parameter, the quasar spectrum exhibits surprisingly little absorption associated with the galaxy pair with a total H\,I column density no greater than $\log N({\rm H\,I})/{\rm cm}^{-2} = 13.65 \pm 0.05$ and no detected heavy element absorption within the virial velocity, $\Delta v = \pm 200$ \kms.

In this section, we summarize the H\,I column density measurement and and limits on the column densities of heavy element ions enabled by the COS and HIRES spectra of \PG.

We measure the absorber properties based on a Voigt profile analysis using the VPFIT package \citep{Carswell:1987}. In cases of component blending, we employ the minimum number of absorption components required to achieve a reasonable $\chi^2$ value. The COS line-spread function (LSF) exhibits broad wings which contain up to $40$\% of the flux \citep{Ghavamian:2009}. In order to properly account for the absorption that falls in the LSF wings, we convolve the Voigt profile models with the empirical, wavelength-dependent COS LSF\footnote{http://www.stsci.edu/hst/cos/performance/spectral\_resolution/}. The continuum normalized COS spectrum and best-fit models are displayed in Figure \ref{figure:absorption}.

\begin{figure}
	\centering
	\includegraphics[scale=0.43]{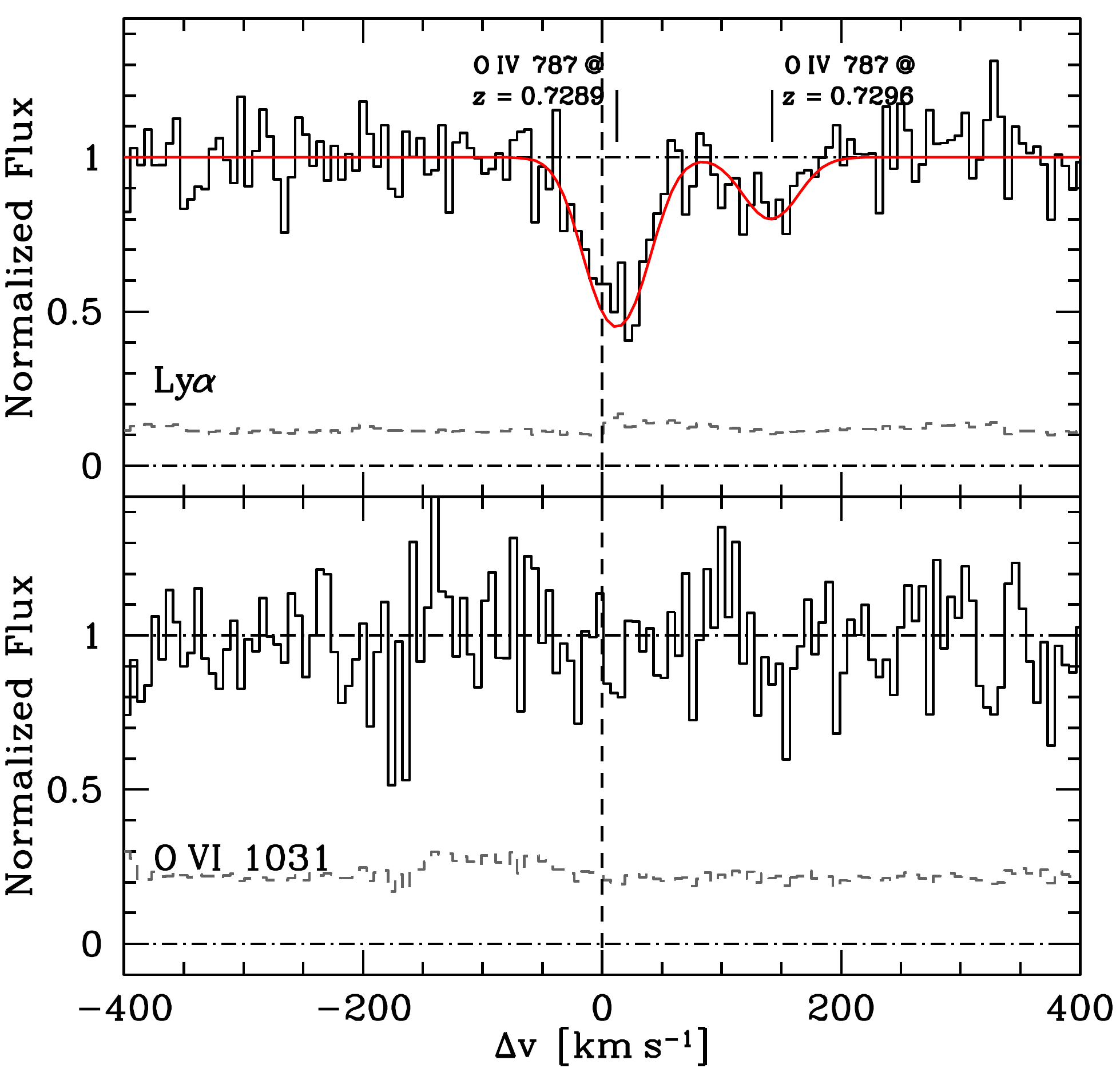}
	\caption{
Same as Figure \ref{figure:absorption} but with contaminating features
that are constrained by other transitions in the COS spectrum removed (see Section 4 for details).
The absorption feature shown in the top panel could be due to \lya\, associated
with the galaxy pair at $z=0.12$ and/or O\,IV $\lambda787$
from the multi-phase gas associated with an LLS at $z = 0.7289$.
Attributing all of the observed absorption to \lya\, at $z=0.12$ leads
to a total H\,I column density of $\log \, N({\rm H\,I})/{\rm cm}^{-2} = 13.65\pm0.05$ (red solid line),
which sets an upper limit to the amount of H\,I that can be associated with the galaxy pair. 
In the bottom panel, we show that no excess absorption is observed after removing
the contaminating features and we place a limit of $\log\,N\rm{(O\,VI)/{\rm cm}^{-2}} < 13.7$
on the OVI associated with the galaxy pair. }
	\label{figure:absorption1}
\end{figure}

The \lya\, absorption associated with the galaxy pair (see the top left panel of Figure \ref{figure:absorption}) is contaminated
by \lyb\, at $z=0.3278$, but the strength of  the contaminating \lyb\, absorption
is constrained by the \lya\, and ${\rm Ly}\gamma$\,
features observed in the COS spectrum.
In order to remove the \lyb\, contamination, we perform a simultaneous Voigt
profile fit to the  \lya\, and ${\rm Ly}\gamma$\, transitions,
resulting in a model for the corresponding \lyb.
We then divide the data by the modeled \lyb\, absorption to remove
the contamination due to the $z=0.3278$ system
(see the top panel of Figure \ref{figure:absorption1}). 

After removal of the \lyb\, absorption, 
the data show significant absorption which could be due to \lya\,
associated with the galaxy pair, O\,IV $\lambda 787$
absorption from a Lyman limit system (LLS) at $z=0.7289$,
or a combination of the two.
\citealt{Lehner:2013} report a metallicity and ionization parameter
limit of ${\rm [X/H]}<-2.0$ and $\log\,U>-3.2$ based
on the H\,I, C\,III, and O\,III column densities detected in the LLS.
With metallicity and ionization parameter within this range,
we do not expect the O\,IV associated with the LLS to
to significantly contaminate the \lya\, associated with the
galaxy pair.
However, O\,VI is detected in the LLS indicating a multi-phase nature,
and the velocity structure seen in the LLS is consistent
with the two absorption components seen in the top panel of Figure \ref{figure:absorption1}.
Consequently, a significant portion of the observed
absorption could be due to O\,IV associated with the LLS.
We therefore place a limit on the possible presence of H\,I
associated with the galaxy pair by attributing all of the observed
absorption to \lya\, at $z=0.12$.
The data are best fit by two Voigt components at 
$\Delta v = +7$ and $+140$ \kms\, 
with column densities of $\log N({\rm H\,I})/{\rm cm}^{-2}=13.5\pm0.06,13.0\pm0.07$
and line widths of $b=28\pm4,28\pm7$ \kms\, respectively.
The line-widths correspond to an upper limit on the gas temperature of $T<5\times10^4$ K.
We note that the line-width measurements are only valid under the assumption
that the observed absorption is largely due to \lya\, associated with the galaxy pair.

The O\,VI $\lambda1031$ absorption associated with G1 and G2
(see the second panel from the bottom on the left of Figure \ref{figure:absorption})
is contaminated by \lyb\, at $z=0.1265$ and O\,III $\lambda 507$ at $z=1.2776$.
In order to remove the contaminating absorption,
we perform a Voigt profile analysis of the \lya\, absorption at $z=0.1265$
and O\,III $\lambda 702$ absorption at $z=1.2776$.
We use these fits to model the contaminating absorption
and divide the data by the model to remove the contamination.
The results are shown in the bottom panel of Figure \ref{figure:absorption1}).
No excess absorption is observed and we place a $3$-$\sigma$
limit of $\log\,N\rm{(O\,VI)/{\rm cm}^{-2}} < 13.7$
on the O\,VI associated with the galaxy pair.

The other transitions shown in Figure \ref{figure:absorption} are not significantly contaminated. For these transitions, we calculate column density limits by measuring the $3$-$\sigma$ equivalent width limit over a $300$ \kms\, window (significantly broader than the wings of the COS LSF) centered at $z=0.1202$ and convert to the corresponding $3$-$\sigma$ limit in column density assuming that the gas is optically thin. We detect no other heavy-element absorption associated with G1 and G2 with column density limits of $\log\,N({\rm Si\,II})/{\rm cm}^{-2} < 12.4$, $\log\,N({\rm C\,II})/{\rm cm}^{-2} < 13.3$, $\log\,N({\rm Si\,III})/{\rm cm}^{-2} < 12.2$, $\log\,N({\rm Mg\,II})/{\rm cm}^{-2} < 12.2$, $\log\,N({\rm Si\,IV})/{\rm cm}^{-2} < 12.7$, and $\log\,N({\rm C\,IV})/{\rm cm}^{-2} < 13.2$. The column density limits of the absorption associated with the galaxy pair are summarized in Table \ref{table:absorption}. We are not able to place a limit on the presence of N\,V because the $\lambda \lambda 1238,1242$ doublet is contaminated by a Lyman limit system at $z=0.5185$.

We performed a series of Cloudy \citep[Version 13; see][]{Ferland:2013} photoionization 
models in an attempt to constrain the physical
conditions of the galaxy halo gas, but the H\,I, O\,VI, and other
heavy element column density limits are not
sufficient to place constraints on the
metallicity and ionization state of the gas.

\section{Discussion and Conclusions} \label{section:discussion}

\begin{figure}
	\centering 
	\includegraphics[scale=0.43]{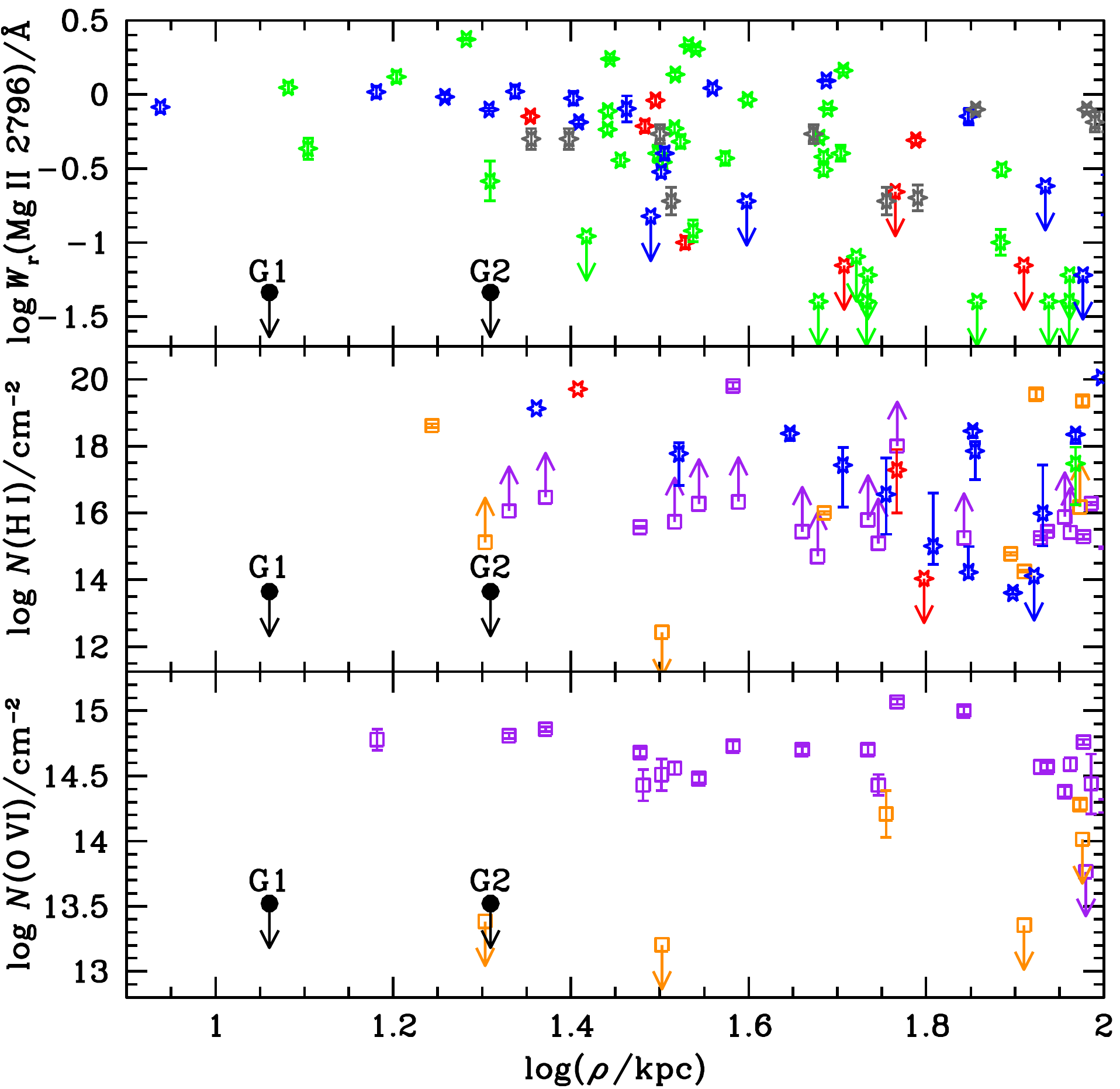} \caption{{\it top}: Equivalent width of Mg\,II $\lambda 2796$ versus $\rho$ reproduced from \protect \cite{Chen:2010} after converting to the cosmology used in this work. The galaxies are shown as six-sided stars and separated into classes by color with S0 galaxies in red, disk galaxies in green, and irregulars in blue (see \protect \citealt{Chen:2010} for details). In addition, group galaxies are shown in gray. Detections are shown with $1$-$\sigma$ uncertainties and non-detections are indicated by arrows showing $2$-$\sigma$ upper limits. G1 and G2 are shown in black and labelled. The Mg\,II equivalent width limit for G1 and G2 falls more than an order-of-magnitude below the mean expected from the \protect \cite{Chen:2010} sample. {\it middle}: H\,I column density as a function of impact parameter reproduced from \protect \cite{Chen:2001a} and \protect \cite{Werk:2013} after converting to the cosmology used in this work. Galaxies from \protect \cite{Chen:2001a} are shown with a six-sided star and $1$-$\sigma$ error bars. Upper limits are $2$-$\sigma$ and are indicated by arrows. The sample is divided into S0 (red), early-type disk (green), and late-type disk (blue) galaxies. Galaxies from \protect \cite{Werk:2013} are shown as open squares. The \protect \cite{Werk:2013} star-forming sample is shown in purple and the passive sample is shown in orange. G1 and G2 are shown as black dots and labelled. The H\,I column density limit of gas associated with G1 and G2 falls more than two orders-of-magnitude below the mean expected from the \protect \cite{Chen:2001a} and \protect \cite{Werk:2013} samples. {\it bottom}: O\,VI column density as a function of impact parameter. Symbols and coloring follow those of the middle panel. The O\,VI column density limit of gas associated with G1 and G2 falls an order-of-magnitude below the mean expected of galaxies with detectable star-formation. } \label{figure:rho} 
\end{figure}

We have identified an interacting pair of galaxies at $z=0.12$ and $\rho<20$ kpc from the sightline toward the background quasar \PG\, at $z_{\rm em}=1.328$. A search in the photometric redshift catalog of SDSS galaxies shows that the galaxy pair may be part of a small galaxy group (as discussed at the end of Section \ref{section:galaxy}). The quasar sightline reveals surprisingly little absorption associated with the galaxy pair with a total H\,I column density no greater than  $\log \, N({\rm H\,I})/{\rm cm}^{-2}=13.65\pm0.05$ and no detected heavy-element absorption. The dearth of absorption is unusual when compared with results from previous absorption-line surveys in the vicinities of field galaxies. In the top panel of Figure \ref{figure:rho}, we show the $W_r({\rm Mg\,II}\,\lambda 2796)-\rho$ relation from \cite{Chen:2010} along with the limit on the equivalent width of Mg\,II $\lambda 2796$ associated with G1 and G2. All 15 field galaxies in the \cite{Chen:2010} sample at $\rho \lesssim 25$ kpc from the background quasar sightline exhibit a strong associated Mg\,II absorber, indicating a unity gas covering fraction. In contrast, the Mg\,II equivalent width limit for G1 and G2 falls more than an order-of-magnitude below the mean expected from the \cite{Chen:2010} sample.

Similarly, we show the H\,I column density versus $\rho$ relation from \cite{Chen:2001a} and \cite{Werk:2013} in the middle panel of Figure \ref{figure:rho} along with the H\,I column density of gas associated with G1 and G2 which falls well below the mean expected from the \cite{Chen:2001a} and \cite{Werk:2013} samples. We estimate that the probability of seeing such little absorption in a sightline at $\rho < 40$ kpc to two non-interacting galaxies is less than $0.2$\%

In addition, we detect no O\,VI absorption associated with the galaxy pair indicating an absence of the warm-hot halo gas commonly seen around galaxies with detectable star-formation \citep[e.g.][]{Chen:2001a, Tumlinson:2011}. In the bottom panel of Figure \ref{figure:rho}, we show the O\,VI column density versus $\rho$ relation from \cite{Werk:2013} along with the O\,VI column density limit of gas associated with G1 and G2. The O\,VI column density limit falls an order-of-magnitude below the mean expected from galaxies with detectable star-formation in the \cite{Werk:2013} sample.

Cool gas is commonly detected in the halos of galaxies out to several hundred kpc in both $21$-cm studies at low redshift \citep[e.g.][]{Oosterloo:2007, Grossi:2009} and in more sensitive absorption-line studies at higher redshift \cite[e.g.][]{Chen:2010, Tumlinson:2011, Thom:2012}. At the same time, a reduced H\,I gas mass is commonly seen in groups of galaxies \citep[e.g.][]{Verdes:2001, Pisano:2011, Catinella:2013}. The gaseous halos of the interacting galaxies presented here could have been tidally stripped, expelled, or heated to $T\gg10^6$ K by an interaction-triggered feedback mechanism such as a starburst or an active-galactic nucleus (AGN) driven wind. However, there is no observational evidence for a recent burst of star-formation or AGN activity, making more direct stripping or heating by the galaxy interaction a more likely explanation for the dearth of absorption near the galaxy pair. The reservoirs of cool gas in the galaxy halos have been significantly depleted possibly indicating that the galaxies are in the process of interaction-induced quenching. Based on a single sightline alone, however, we cannot rule out the presence of significant cool gas in the galaxy halos since the halos could be characterized by a reduced but non-zero covering fraction. Indeed, our observations cannot rule out the presence of a substantial H\,I gas disk. For galaxies with stellar masses similar to that of G1, we expect an H\,I gas disk with mass of $M_{\rm H\,I} \sim 10^{9-10}\, M_\odot$ based on the stellar to H\,I mass relation from \cite{Catinella:2013}. Adopting the H\,I disk mass to cross-section scaling relation from \cite{Rosenberg:2003}, we infer an expected H\,I disk radius of $R_{\rm H\,I} \sim 7-30$ kpc. It is therefore possible that G1 possesses an H\,I disk within the normal mass range but which is not large enough to intercept the sightline of \PG.

The transparent sightline at low impact parameter to two interacting galaxies presented here underscores the potential impact of galaxy interactions on the gaseous halos of galaxies. It suggests that galaxy interactions can significantly reduce the reservoir of cool gas in galaxy halos without a strong feedback mechanism. However, the relationship between galaxy environment and halo gas content remains largely unexplored. We are aware of only two other low impact parameter sightlines probing close and possibly interacting galaxy pairs. One consists of two red ($M_B - M_R > 1.0$) galaxies at $z=0.226$ separated by $17$ kpc and probed by a quasar sightline at projected distances of $\rho = 30$ kpc. \cite{Chen:2010} report the detection of an associated Mg\,II absorption system at $\Delta v = 0$ \kms\, with $W_r({\rm Mg\,II\,\lambda 2796}) = 0.54\pm0.08\,$\AA\, for this galaxy pair. The other galaxy pair consists of two star-forming galaxies separated by $16$ kpc at $z=0.095$ and is probed at $\rho < 30$ kpc by the sightline of the UV-bright quasar SDSS J0820+2334 \citep[see ][]{Werk:2012}. There is a moderate \lya\, absorber with $\log\,N({\rm H\,I})/{\rm cm}^{-2} = 14.0\pm0.1$ associated with this galaxy pair ($\Delta v = 0$) with no detected metal-line absorption.

The observed H\,I absorption is significantly weaker than what is expected for a sightline at low impact parameter to non-interacting galaxies. There is a strong \lya\ system with $\log\,N({\rm H\,I})/{\rm cm}^{-2} = 15.3 \pm 0.5$ and associated C\,IV and O\,VI absorption at $\Delta v = +300$ \kms, the same velocity as a more luminous, star-forming galaxy at $\rho = 140$ kpc. The presence the more luminous, star-forming galaxy complicates the interpretation of this system. Further exploration of the impact of galaxy environment and interactions on the gaseous content of galaxy halos will likely require a systematic survey of quasar sightlines probing the environments of group galaxies.

\section*{Acknowledgements} We are grateful to M. Rauch and W. Sargent for obtaining deep optical images of the \PG\, field for us.
We thank the anonymous referee for highly constructive comments and feedback.
We are grateful for the support provided by the staff at the Las Campanas Observatories. This paper includes data gathered with the 6.5 meter Magellan Telescopes located at Las Campanas Observatory, Chile.
Some of the data presented herein were obtained at the W. M. Keck Observatory, which is operated as a scientific partnership among the California Institute of Technology, the University of California and the National Aeronautics and Space Administration. The Observatory was made possible by the generous financial support of the W. M. Keck Foundation.
SDJ acknowledges support from a National Science Foundation Graduate Research Fellowship.
Partial support for this research was provided by NASA through grant HST-GO-11741.
This research has made use of the NASA Astrophysics Data System (ADS) and the NASA/IPAC Extragalactic Database (NED) which is operated by the Jet Propulsion Laboratory, California Institute of Technology, under contract with the National Aeronautics and Space Administration. \footnotesize{ 
\bibliographystyle{mn} 
\bibliography{manuscript}}

\bsp \label{lastpage}

\end{document}